\documentclass[sigconf]{acmart}
\AtBeginDocument{%
  }

\copyrightyear{2026}
\acmYear{2026}
\setcopyright{cc}
\setcctype{by}
\acmConference[CSCW Companion '26]{Companion of the Computer-Supported Cooperative Work and Social Computing}{October 10--14, 2026}{Salt Lake City, UT, USA}
\acmBooktitle{Companion of the Computer-Supported Cooperative Work and Social Computing (CSCW Companion '26), October 10--14, 2026, Salt Lake City, UT, USA}
\acmDOI{10.1145/3785651.3831419}
\acmISBN{979-8-4007-2378-0/2026/10}

\begin{document}

\title[CoBranchMR]{CoBranchMR: Supporting Parallel Design and Conflict Resolution in Mixed Reality}

\newcommand{\toolname}{\textsc{\textbf{CoBranchMR}}}

\author{Niloofar Sayadi}
\affiliation{%
  \institution{University of Notre Dame}
  \department{Computer Science and Engineering}
  \city{Notre Dame}
  \state{IN}
  \country{USA}}
\email{nsayadi2@nd.edu}

\author{Kaiyuan Tang}
\affiliation{%
  \institution{University of Notre Dame}
  \department{Computer Science and Engineering}
  \city{Notre Dame}
  \state{IN}
  \country{USA}}
\email{ktang2@nd.edu}

\author{Yunhao Xing}
\affiliation{%
  \institution{University of Notre Dame}
  \department{Computer Science and Engineering}
  \city{Notre Dame}
  \state{IN}
  \country{USA}}
\email{xyunhao@nd.edu}

\author{Simret Gebreegziabher}
\affiliation{%
  \institution{University of Notre Dame}
  \department{Computer Science and Engineering}
  \city{Notre Dame}
  \state{IN}
  \country{USA}}
\email{sgebreeg@nd.edu}

\author{Chaoli Wang}
\affiliation{%
  \institution{University of Notre Dame}
  \department{Computer Science and Engineering}
  \city{Notre Dame}
  \state{IN}
  \country{USA}
}
\email{chaoli.wang@nd.edu}

\author{Diego Gómez-Zará}
\affiliation{%
  \institution{University of Notre Dame}
  \department{Computer Science and Engineering}
  \city{Notre Dame}
  \state{IN}
  \country{USA}
}
\email{dgomezara@nd.edu}

\renewcommand{\shortauthors}{Sayadi et al.}

\begin{abstract}
We present \toolname{}, a mixed reality (MR) system that enables distributed collaborators to work in parallel from different locations on the same digital representation of a physical object. \toolname{} lets users branch an object into editable virtual copies, customize them independently, and then merge their work back into a shared object. When merging copies, the system displays potential conflicts on the object's surface and provides several resolution options. By adopting branch-and-merge workflows for embodied spatial collaboration, \toolname{} introduces a new collaborative interaction model that supports parallel design, conflict resolution, and negotiation in remote creative work.
\end{abstract}

\ccsdesc[500]{Human-centered computing~Mixed / augmented reality}
\ccsdesc[300]{Human-centered computing~Collaborative interaction}
\ccsdesc[300]{Human-centered computing~Computer supported cooperative work}

\keywords{Mixed Reality; Collaborative Design; Branch-and-Merge; Conflict Resolution; 3D Painting; Gaussian Splatting}


\begin{teaserfigure}
    \centering
    \includegraphics[width=\linewidth]{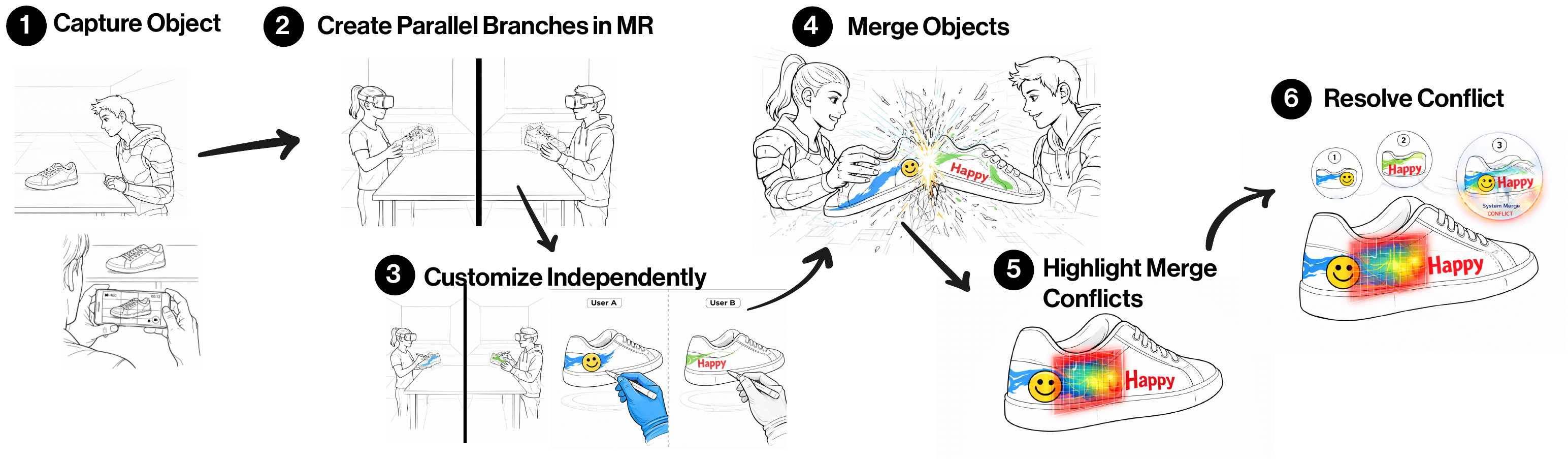}
    \Description{A horizontal workflow diagram with six labeled stages. Left: a hand holds a phone to capture a real sneaker, while two people in MR headsets share a session. Center-left: the two users separate and each customizes their own version of the shoe independently — one adds a smiley face sticker, the other adds the word Happy. Center: the two users physically bring their versions together in a merge attempt, shown with a spark effect. Center-right: the merged shoe is shown with a heatmap overlay highlighting regions where the two edits spatially conflict. Right: a hand selects among circular thumbnail options representing different conflict resolution choices for the overlapping regions.}
    \caption{From capture to conflict resolution: \toolname{} lets collaborators branch a shared 3D object, customize in parallel, and merge their edits, highlighting spatial conflicts and supporting hands-on resolution.}
    \label{fig:teaser-figure}
\end{teaserfigure}

\maketitle

\begin{figure*}[!ht]
  \centering
  \includegraphics[width=1\textwidth]{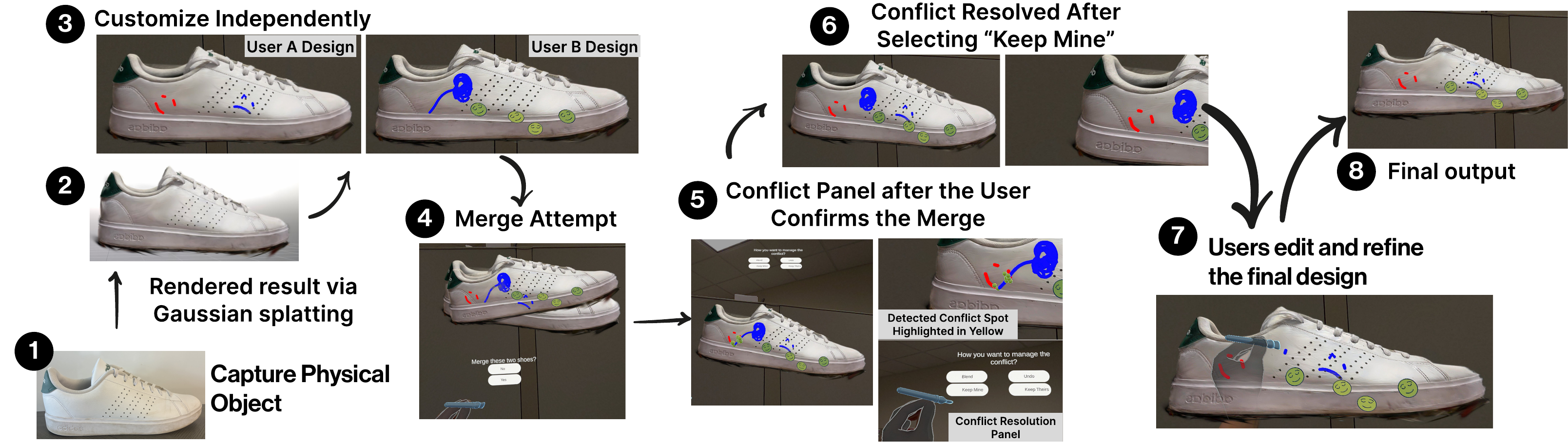}
  \Description{A horizontal system walkthrough diagram with seven stages. Bottom-left: a photograph of a real white sneaker labeled ground-truth shoe image, with an arrow pointing up to a Gaussian splatting reconstruction of the same shoe. Top-left: two side-by-side shoe renders labeled User A Design and User B Design, each showing different painted strokes on the shoe. Center-left: a Merge Attempt stage showing the physical shoe with a glowing blue orb indicating an interaction. Center: a Conflict Panel displayed after the user confirms the merge, showing the shoe with a conflict region highlighted in yellow and a resolution panel with button options below. Top-center-right: two shoe thumbnails labeled Conflict Resolved After Selecting Keep Mine, showing one user's design preserved. Bottom-center-right: a shoe with additional brush strokes, labeled Users editing and refining the final design. Far right: a final rendered shoe photograph labeled Final output.}
  \caption{\toolname{}'s walkthrough: a real shoe is reconstructed via Gaussian splatting, two users independently customize separate branches, a merge attempt triggers conflict detection, and users resolve conflicts before collaboratively refining the final design.}
  \label{Fig2}
\end{figure*}

\section{Introduction}
Collaborative work around artifacts often requires people to coordinate how their individual contributions should be aligned, combined, and reconciled. CSCW research has studied how socio-technical systems can support the additional efforts required to articulate individual actions into a shared single object \cite{strauss1988articulation,schmidt1992taking,sayadi2024feeling}. For distributed collaborators, this articulation work becomes harder because physical objects are situated in only one place, are not easily viewable or editable by multiple users, and require users to establish common ground \cite{bjorn2014does,garg2025remotely}.

Prior CSCW and MR systems have supported remote collaboration around digital workspaces by providing shared virtual context, spatial awareness, and opportunities for joint manipulation \cite{billinghurst1999collaborative,praveena2023periscope,villanueva2022colabar}. These systems are valuable for helping collaborators see and discuss the same workspace. Still, they do not allow them to explore different possibilities in parallel before deciding what to preserve, combine, or discard. Requiring all edits to happen on a shared object can force premature negotiation and limit individual exploration, making it difficult to compare alternative designs.

Version control addresses this coordination problem by allowing collaborators to branch from a shared source, work independently, and merge their changes later. When edits overlap, the system highlights conflicts and requires collaborators to resolve them through negotiation. Recent work has applied version-control concepts to immersive and visual content. VRGit supports collaborative editing of VR scenes \cite{zhang2023vrgit}, while non-linear revision control supports resolving overlapping edits in images \cite{chen2011nonlinear}. However, less is known about how branch-and-merge workflows support distributed collaboration around digital artifacts represented in MR, where conflicts are spatial, embodied, and tied to the surface of an object.

We present \toolname{}, an MR system that supports parallel design and conflict resolution around a shared digital artifact. Figure \ref{fig:teaser-figure} illustrates its branch-and-merge interaction model. It employs 3D Gaussian splatting \cite{Kerbl2023} to reconstruct a photorealistic virtual representation of a physical object, which collaborators can branch into separate editable virtual copies. Users customize these copies independently and then merge their work back into a single shared version through an embodied interaction in mixed reality. When collaborators make overlapping edits, the system visualizes the conflict directly on the object surface. It provides options to keep one version, keep the other, blend both, or undo the merge.

Through this demo, we show how branch-and-merge interaction can support a collaborative process that moves between individual exploration and joint decision-making. \toolname{} contributes to CSCW by demonstrating a novel interaction model for coordinating parallel work, maintaining conflict awareness, and negotiating shared outcomes in distributed creative work.
\section{System Description}
Figure \ref{Fig2} provides an overview of the \toolname{} workflow. The system first reconstructs a 3D Gaussian splat representation of a physical object. Two distributed users then create separate branches, customize their copies independently, and bring them together to initiate a merge. \toolname{} detects overlapping edits, displays the conflicts on the object, and provides resolution options before users continue refining the merged design collaboratively.

The system runs on two MR headsets connected over a local network for real-time multiplayer synchronization. A physical object is captured once using 3D Gaussian splatting and imported into the system as a photorealistic object. An invisible AI-generated proxy mesh is overlaid on the splat to support physics, collision detection, and painting interactions.

\subsubsection*{Branch: Independent Painting}
Once the physical object is in the environment, each user uses the virtual brush to press a button that spawns their own copy of the object. The object can be grabbed, moved, and rotated freely using hand tracking. To paint, the user dips the brush into a colored cube and then draws on the object surface. To erase, they dip the brush into a white eraser.

The painting system creates smooth, continuous strokes directly on the object surface. A raycast fires from the brush tip in all six directions to find the exact surface contact point. Strokes are stored in the object's local space, so they stay attached to the object when it is moved. The eraser can remove parts of strokes by splitting them at the erased section, similar to how erasing works in drawing apps. Each stroke is labeled with the user's ID (\texttt{User A} or \texttt{User B}) when it is created. This identity tag is used later by the conflict detection system.

\subsubsection*{Merge: Bringing Copies Together}
When both users agree on their designs, they physically carry their object copies toward each other in MR space until the copies touch. A confirmation panel appears in front of the user asking: ``\textit{Merge these two objects?}'' If they choose \textit{``Yes,''} all strokes from one object are transferred to the other using coordinate-space conversion to keep them in the correct position. The other copy is then removed from the scene.

\subsubsection*{Conflict Detection}
After the merge, a Conflict Manager script scans the combined object. It separates the strokes into two lists based on their user ID labels. It then compares every point in User A's strokes against every point in User B's strokes using 3D distance $d = \lVert \mathbf{p}_A - \mathbf{p}_B \rVert$, where $\mathbf{p}_A$ and $\mathbf{p}_B$ represent stroke points.


If the distance between any two points from different users is less than 5 cm, a conflict is flagged. For each conflict found, a small, glowing yellow sphere is placed at that location on the object's surface so both users can immediately see where their designs overlap. The system uses short-circuiting: once a conflict is found between a pair of strokes, it stops checking the remaining points in that pair to keep the frame rate smooth in MR.

\subsubsection*{Conflict Resolution}
When conflicts are found, a resolution panel appears in the user's view. The panel rotates to always face the user (billboard behavior), offering four choices: (a) \textit{``Keep Mine,''} which keeps the local user's painting and removes the other user's work; (b) \textit{``Keep Theirs,''} which removes the local user's strokes and keeps the other user's painting; (c) \textit{``Blend,''} which keeps all strokes and leaves both sets of paint visible together; and (d) \textit{``Undo,''} which reverses the merge and restores each user’s separate copy.

\subsubsection*{Implementation}
The system is built on Unity 6 \cite{unity6_software} and runs on Meta Quest 3 headsets. Key components include Unity-VR-Gaussian-Splatting \cite{ninjamode2023}, a Unity package that enables Gaussian splats to be rendered in VR/MR environments; an AI-generated proxy mesh created with Meshy AI \cite{meshyai} for collision detection, painting interactions, and physics; Normcore (Normal.Realtime) \cite{normcore2019} for real-time multiplayer sync and networked object spawning; Meta XR Building Blocks \cite{metaxrbb2023} for hand tracking, grabbing, and UI interaction; and Unity LineRenderer \cite{unity_linerenderer} for smooth stroke painting on 3D surfaces.
\section{Future Work}
Future work will expand \toolname{} in four directions. First, we plan a controlled dyad study to examine how collaborators negotiate conflicts in 3D creative work, including their coordination strategies, negotiation patterns, and satisfaction with merged outcomes. Second, we will study how decisions made on digital representations transfer to physical objects, including whether users feel confident implementing merged designs on the real artifact and whether the digital representation is sufficiently faithful to support that transition. Third, we will improve the manipulation of Gaussian splats. The current system supports painting on an AI-generated proxy mesh overlaid on the splat, which can cause slight stroke offsets when the two are misaligned. Direct runtime editing of splat colors would reduce this issue and improve editing and export workflows. Finally, future versions could support more users and objects, including pairwise or group merges from a shared base object, as well as generative AI features to recommend related virtual objects during ideation.

\section{Conclusion}
We presented \toolname{}, an MR system that adapts branch-and-merge workflows to collaborative 3D object customization. \toolname{} enables distributed users to edit parallel versions of a digital object and resolve surface-level conflicts by selecting either version, blending both versions, or undoing the merge. This interaction model extends CSCW research on coordination, awareness, shared artifacts, and conflict negotiation into embodied MR settings. The prototype opens avenues for studying remote creative work involving digital representations of physical artifacts.

\section*{AI Disclosure}
Figure 1 was created by the first author and includes AI-generated illustrative elements using Google Gemini to depict the CoBranchMR workflow. Figure 2 was created by the authors from the CoBranchMR prototype. AI-based writing tools were used to support grammar, clarity, and language revision. All technical content, claims, and final text were reviewed and approved by the authors.

\begin{acks}
    This work was supported by the Alfred P. Sloan Foundation (Grant \#G-2024-22427).
\end{acks}

\bibliographystyle{ACM-Reference-Format}

\bibliography{main}

@inproceedings{zhang2023vrgit,
    author = {Zhang, Lei and Agrawal, Ashutosh and Oney, Steve and Guo, Anhong},
    title = {VRGit: A Version Control System for Collaborative Content Creation in Virtual Reality},
    year = {2023},
    isbn = {9781450394215},
    publisher = {Association for Computing Machinery},
    address = {New York, NY, USA},
    doi = {10.1145/3544548.3581136},
    booktitle = {Proceedings of the 2023 CHI Conference on Human Factors in Computing Systems},
    articleno = {36},
    numpages = {14},
    location = {Hamburg, Germany},
    series = {CHI '23}
}

@article{chen2011nonlinear,
  title={Nonlinear revision control for images},
  author={Chen, Hsiang-Ting and Wei, Li-Yi and Chang, Chun-Fa},
  journal={ACM Transactions on Graphics (TOG)},
  volume={30},
  number={4},
  pages={1--10},
  year={2011},
  publisher={ACM New York, NY, USA}
}

@article{sayadi2024feeling,
    author = {Sayadi, Niloofar and Co, Sadie and G\'{o}mez-Zar\'{a}, Diego},
    title = {"Feeling that I was Collaborating with Them:" A 20-year Scoping Review of Social Virtual Reality Leveraging Collaboration},
    year = {2026},
    issue_date = {April 2026},
    publisher = {Association for Computing Machinery},
    address = {New York, NY, USA},
    volume = {10},
    number = {2},
    doi = {10.1145/3788053},
    journal = {Proc. ACM Hum.-Comput. Interact.},
    month = may,
    articleno = {CSCW017},
    numpages = {33}
}

@misc{ninjamode2023,
  author = {ninjamode},
  title = {Unity-VR-Gaussian-Splatting},
  year = {2023},
  publisher = {GitHub},
  license = {MIT License},
  howpublished = {\url{https://github.com/ninjamode/Unity-VR-Gaussian-Splatting}},
  note = {Accessed: 2026}
}

@misc{meshyai,
  author = {{Meshy AI}},
  title = {Meshy: AI 3D Model Generator},
  howpublished = {\url{https://www.meshy.ai}},
  note = {Accessed: May 2026}
}

@misc{normcore2019,
  author = {{Normal}},
  title = {Normcore: Multiplayer SDK for Unity},
  year = {2019},
  publisher = {Normal},
  howpublished = {\url{https://normcore.io}},
  note = {Accessed: 2026}
}

@misc{metaxrbb2023,
  author = {{Meta Platforms}},
  title = {Meta {XR} Building Blocks for Unity},
  year = {2023},
  publisher = {Meta Platforms},
  howpublished = {\url{https://developers.meta.com/horizon/documentation/unity/bb-overview/}},
  note = {Accessed: 2026}
}

@misc{unity_linerenderer,
  author = {{Unity Technologies}},
  title = {Line Renderer --- {Unity} Documentation},
  year = {2024},
  publisher = {Unity Technologies},
  howpublished = {\url{https://docs.unity3d.com/Manual/class-LineRenderer.html}},
  note = {Accessed: 2026}
}

@article{praveena2023periscope,
  title={Periscope: A robotic camera system to support remote physical collaboration},
  author={Praveena, Pragathi and Wang, Yeping and Senft, Emmanuel and Gleicher, Michael and Mutlu, Bilge},
  journal={Proceedings of the ACM on Human-Computer Interaction},
  volume={7},
  number={CSCW2},
  pages={1--39},
  year={2023},
  publisher={ACM New York, NY, USA}
}

@article{villanueva2022colabar,
  title={Colabar: A toolkit for remote collaboration in tangible augmented reality laboratories},
  author={Villanueva, Ana and Zhu, Zhengzhe and Liu, Ziyi and Wang, Feiyang and Chidambaram, Subramanian and Ramani, Karthik},
  journal={Proceedings of the ACM on Human-Computer Interaction},
  volume={6},
  number={CSCW1},
  pages={1--22},
  year={2022},
  publisher={ACM New York, NY, USA}
}

@inproceedings{billinghurst1999collaborative,
  title={Collaborative mixed reality},
  author={Billinghurst, Mark and Kato, Hirokazu},
  booktitle={Proceedings of the first international symposium on mixed reality},
  volume={1},
  year={1999}
}

@article{schmidt1992taking,
  title={Taking CSCW seriously: Supporting articulation work},
  author={Schmidt, Kjeld and Bannon, Liam},
  journal={Computer supported cooperative work (CSCW)},
  volume={1},
  number={1},
  pages={7--40},
  year={1992},
  publisher={Springer}
}

@article{strauss1988articulation,
  title={The articulation of project work: An organizational process},
  author={Strauss, Anselm},
  journal={Sociological Quarterly},
  volume={29},
  number={2},
  pages={163--178},
  year={1988},
  publisher={Wiley Online Library}
}

@article{bjorn2014does,
  title={Does distance still matter? Revisiting the CSCW fundamentals on distributed collaboration},
  author={Bj{\o}rn, Pernille and Esbensen, Morten and Jensen, Rasmus Eskild and Matthiesen, Stina},
  journal={ACM Transactions on Computer-Human Interaction (TOCHI)},
  volume={21},
  number={5},
  pages={1--26},
  year={2014},
  publisher={ACM New York, NY, USA}
}

@article{garg2025remotely,
  title={What Remotely Matters? Understanding Individual, Team, and Organizational Factors in Remote Work at Scale},
  author={Garg, Kapil and G{\'o}mez-Zar{\'a}, Diego and Gerber, Elizabeth and Gergle, Darren and Contractor, Noshir and Massimi, Michael},
  journal={Proceedings of the ACM on Human-Computer Interaction},
  volume={9},
  number={7},
  pages={1--47},
  year={2025},
  publisher={ACM New York, NY, USA}
}

@software{unity6_software,
  author       = {{Unity Technologies}},
  title        = {Unity},
  version      = {6},
  year         = {2024},
  url          = {https://unity.com/},
  organization = {Unity Technologies},
}

@article{Kerbl2023,
author = {Kerbl, Bernhard and Kopanas, Georgios and Leimkuehler, Thomas and Drettakis, George},
title = {3D Gaussian Splatting for Real-Time Radiance Field Rendering},
year = {2023},
issue_date = {August 2023},
publisher = {Association for Computing Machinery},
address = {New York, NY, USA},
volume = {42},
number = {4},
issn = {0730-0301},
url = {https://doi.org/10.1145/3592433},
doi = {10.1145/3592433},
journal = {ACM Trans. Graph.},
month = jul,
articleno = {139},
numpages = {14}
}

\end{document}